\documentclass[twoside,twocolumn,9pt]{article}
\usepackage{extsizes}
\usepackage[super,sort&compress,comma]{natbib} 
\usepackage[version=3]{mhchem}
\usepackage[left=1.5cm, right=1.5cm, top=1.785cm, bottom=2.0cm]{geometry}
\usepackage{balance}
\usepackage{times,mathptmx}
\usepackage{sectsty}
\usepackage{graphicx} 
\usepackage{lastpage}
\usepackage[format=plain,justification=justified,singlelinecheck=false,font={stretch=1.125,small,sf},labelfont=bf,labelsep=space]{caption}
\usepackage{float}
\usepackage{fancyhdr}
\usepackage{fnpos}
\usepackage[english]{babel}
\addto{\captionsenglish}{%
  
}
\usepackage{array}
\usepackage{droidsans}
\usepackage{charter}
\usepackage[T1]{fontenc}
\usepackage[usenames,dvipsnames]{xcolor}
\usepackage{setspace}
\usepackage[compact]{titlesec}
\usepackage{hyperref}

\usepackage{epstopdf}

\definecolor{cream}{RGB}{222,217,201}

\begin{document}
\pagestyle{fancy}
\thispagestyle{plain}
\fancypagestyle{plain}{

\renewcommand{\headrulewidth}{0pt}
}
\makeatletter 
\newlength{\figrulesep} 
\setlength{\figrulesep}{0.5\textfloatsep} 

\newcommand{\topfigrule}{\vspace*{-1pt}%
\noindent{\color{cream}\rule[-\figrulesep]{\columnwidth}{1.5pt}} }

\newcommand{\botfigrule}{\vspace*{-2pt}%
\noindent{\color{cream}\rule[\figrulesep]{\columnwidth}{1.5pt}} }

\newcommand{\dblfigrule}{\vspace*{-1pt}%
\noindent{\color{cream}\rule[-\figrulesep]{\textwidth}{1.5pt}} }

\makeatother

\twocolumn[
 \begin{@twocolumnfalse}
\sffamily
\begin{tabular}{m{1.5cm} p{13.5cm} }

&\noindent\LARGE{\textbf{Flow and fracture near the sol-gel transition of silica nanoparticle suspensions$^\dag$}} \\
\vspace{0.3cm} & \vspace{0.3cm} \\

 & \noindent\large{Gustavo E. Gimenes\textit{$^{a}$} and Elisabeth Bouchaud\textit{$^b$} } \\

 &\noindent\normalsize{We analyze the evolution of the mechanical response of a colloidal suspension to an external tensile stress, from fracture to flow, as a function of the distance from the sol-gel transition. We cease to observe cracks at a finite distance from the transition. In an intermediate region where the phenomenon is clearly hysteretic, we observe the coexistence of both flow and fracture. Even when cracks are observed, the material in fact flows over a distance that increases in the vicinity of the transition. 
} \\

\end{tabular}

 \end{@twocolumnfalse} \vspace{0.6cm}
  ]


\renewcommand*\rmdefault{bch}\normalfont\upshape
\rmfamily
\section*{}
\vspace{-1cm}


\footnotetext{\textit{$^{a}$~PSL Research University, Institut Pierre-Gilles de Gennes, ESPCI, UMR Gulliver, 8 rue Jean Calvin, 75005 Paris, France. Tel: +33 685990809; E-mail: gustavo.gimenes@espci.fr}}
\footnotetext{\textit{$^{b}$~PSL Research University, Institut Pierre-Gilles de Gennes, ESPCI, UMR Gulliver, 8 rue Jean Calvin, 75005 Paris, France, and Universit\'e Paris-Saclay, CEA-Saclay, SPEC, 91191 Gif-sur-Yvette Cedex, France. }}

\footnotetext{\dag~Electronic Supplementary Information (ESI) available: [Video 1 and 2 show the propagation of cracks at $V$=10$^{-5}$m.s$^{-1}$, for salt concentrations 205 and 198 mmol.L$^{-1}$. Video 3 shows the progression of a viscous front ( in the hysteretic region, salt concentration 202 mmol.L$^{-1}$) which forms a cusp, and subsequently two cracks.]. Video 4 shows the presence of flow at the crack tip for a salt concentration of 196 mmol.L$^{-1}$ See DOI: 10.1039/b000000x/}


\section{Introduction}

		The mechanical failure of soft materials has been an area of increasing developments over the last years, particularly motivated by progress in the understanding and engineering of soft but resilient and tough polymer hydrogels~\cite{Creton_2017,Zhao_2014, Zhang16}. Colloids constitute another soft system which may undergo high stresses in a number of practical applications, ranging from ceramics processing and coatings to food and personal care products~\cite{Zack06}. 
		
		Dilute suspensions of colloidal particles exhibit a liquid-like behavior, and flow under shear~\cite{Zaccarelli:07}. If the stability of the suspension is disturbed, the particles start forming larger and larger clusters. Eventually, the viscosity of the system diverges, and a non-zero Young's modulus emerges. Under loading, the material deforms and may break like a solid. This sol-gel transition is a second-order phase transition, which can be viewed as the formation of a percolating network of aggregated particles~\cite{Djabourov88, Zaccarelli:07}. 
		
		Colloidal gels have a complex mechanical behavior, which can exhibit non-linear phenomena such as the rise of a yielding threshold, observed both in experiments~\cite{Walls_JRheol03} and in simulations~\cite{Roy_2016a, Roy_2016b, Johnson:2018aa, Colombo_2014}, two-step yielding~\cite{Brunel_2016}, creep and delayed failure~\cite{Aime_2018,Lindstr_m_2012}, and strain hardening~\cite{Gisler99,Colombo_2014}. 
		
		The central point of this paper is indeed to analyze the mechanical response of the material to an external tensile stress in the vicinity of the sol-gel transition. This is directly related to the more general question of how does the failure of a solid change close to a rigidity transition, which was considered from a theoretical point of view and on model metamaterials by Vitelli and collaborators~\cite{Driscoll_PNAS16}. The present paper is, to our knowledge, the first experimental attempt to investigate the phenomenon quantitatively in colloids.

	Our system is an aqueous suspension of colloidal silica particles the sol-gel transition of which can be controlled by adding a certain amount of salt~\cite{Cao:10, Giuse:12}. We use a microfluidic device to apply a tensile stress to this extremely soft material. We observe the resulting cracks or viscous fronts propagating through the sample at two different velocities using optical microscopy. We analyze the shapes of the cracks and the displacement fields in the vicinity of their tips. By projecting the measured displacements onto an elastic basis, we obtain a precise description of the structure of the cracks and estimate the energies involved in the process. Results are discussed in the last section.

\section{Materials and Methods}
		 		 
	The chosen material is an aqueous monodisperse solution of silica nanoparticles (Ludox\textsuperscript{\tiny\textregistered} TM-50, Sigma-Aldrich) with a mean diameter of 22 nm, as reported by the manufacturer and confirmed in the literature~\cite{Orts_Gil_2010}.
	The initial particle concentration is 50 wt$\%$. To achieve the desired particle volume fraction of $25.8\%$ and induce the onset of agglomeration, a pre-solution consisting of sodium chloride, to which we add a mixture of carbon black and India ink, is mixed with the silica dispersion. The quantity of NaCl added is adjusted as a function of the desired final ionic strength of the gel. Experiments are performed in the range of added salt concentrations from 190 mmol.L$^{-1}$ to 205 mmol.L$^{-1}$. The final pH of every suspension is 8.5 (as measured with a Mettler-Toledo SevenCompactS220\textsuperscript{\tiny\textregistered} pHmeter). Immediately after mixing, the system is degassed by creating a light vacuum in a syringe for 3 minutes. 
	
\begin{figure*}
	\centering
	{\includegraphics[width=0.275\textwidth]{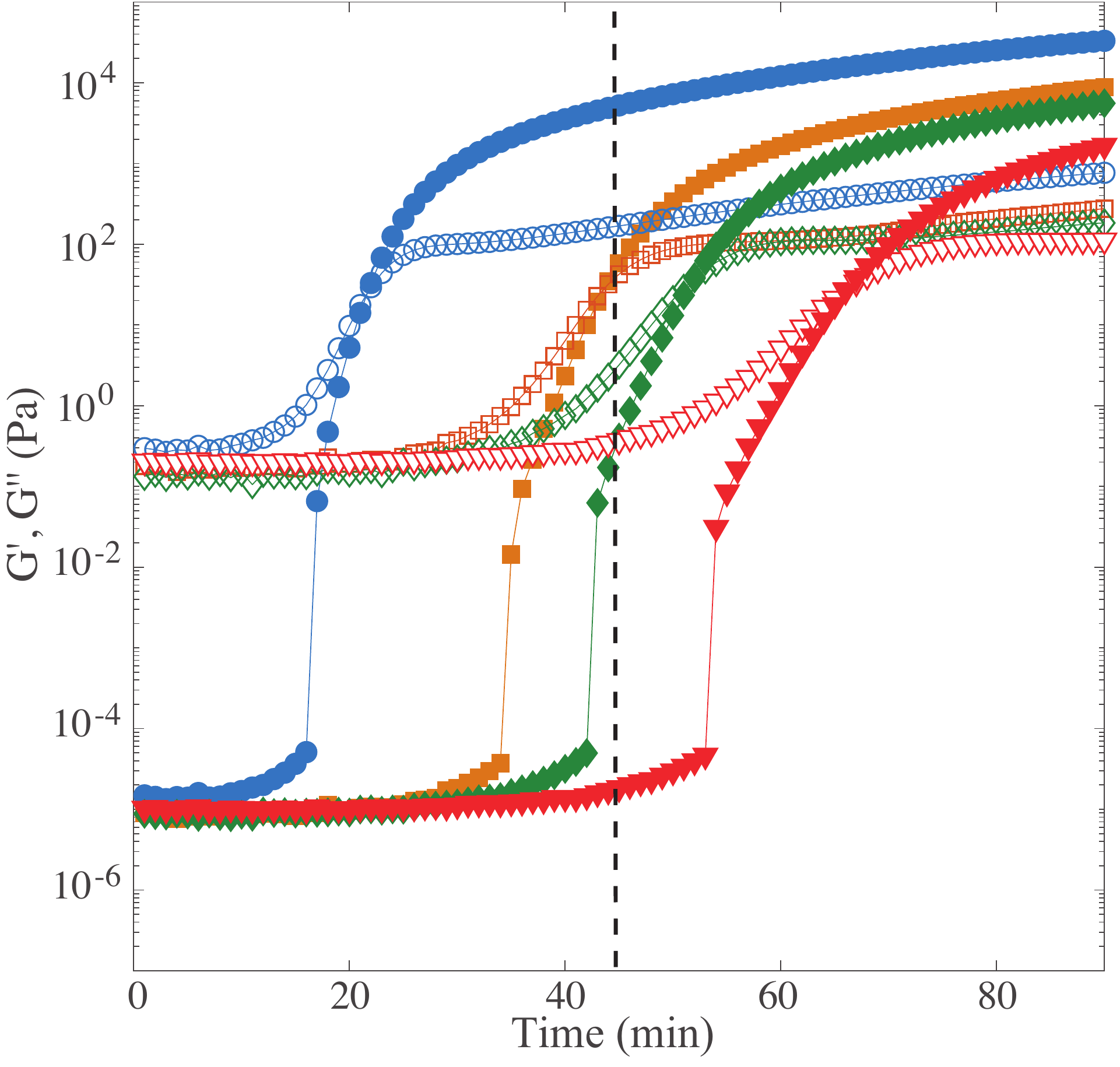}}
	{\includegraphics[width=0.27\textwidth]{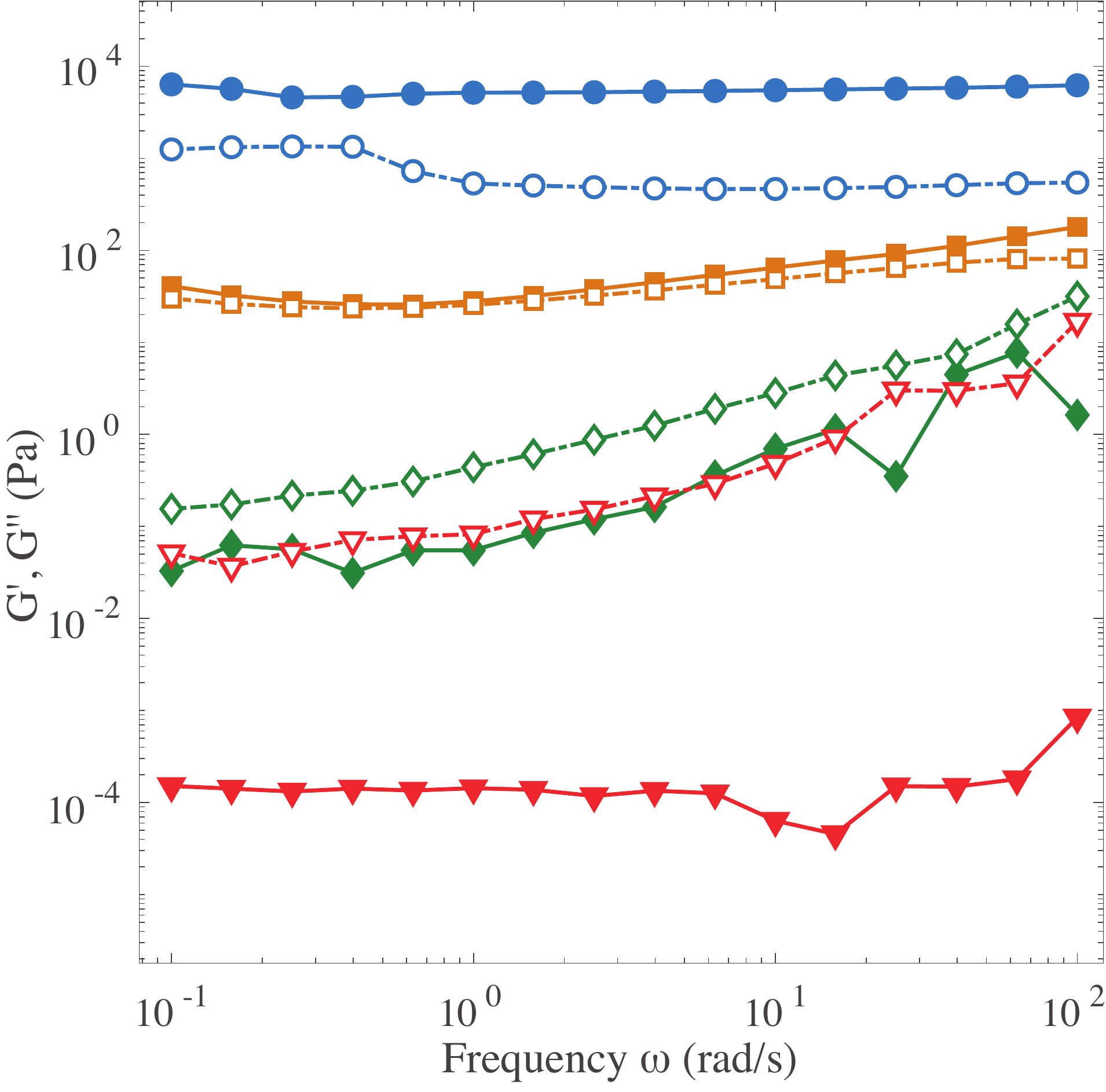}}
	{\includegraphics[width=0.28\textwidth]{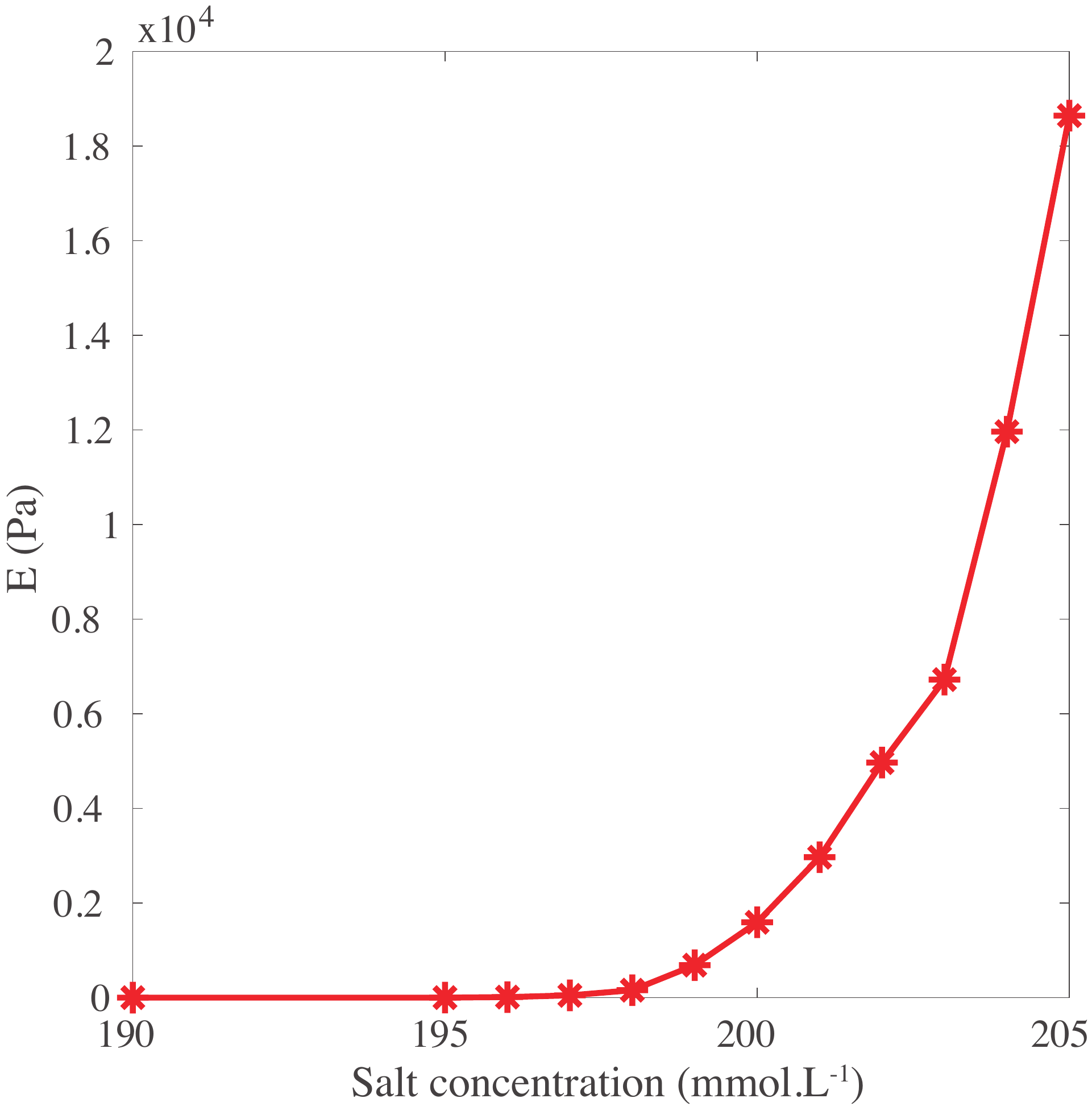}}
	\caption{{Linear rheology measurements. $G"$ (empty symbols) and $G'$ (full symbols). {\bf Left} as a function of time and {\bf Center}  as a function of frequency: red triangles 190 mmol.L$^{-1}$; green rhombus 195 mmol.L$^{-1}$; orange squares 198 mmol.L$^{-1}$; blue circles 205 mmol.L$^{-1}$. The black dotted line on the {\it left} figure indicates that all experiments are performed after a waiting time of 45 minutes. {\bf Right} Young's modulus $E$ as a function of salt concentration. Note that $E$ starts growing beyond 190 mmol.L$^{-1}$.}}
\label{figRheo}
\end{figure*}

	Linear rheology measurements are performed in order to characterize the evolution of the mechanical properties of our gels with the salt concentration. All rheology measurements are made in an MCR 302 Stress-Controlled Rheometer (Anton Paar). The sol-gel transition was monitored by measuring continuously the values of the storage ($G'$) and loss ($G"$) moduli in oscillatory shear flow at constant frequency $\omega=1$ rad/s and fixed strain $0.01$ at the waiting temperature of $50~^\circ$ C. Figure \ref{figRheo}- Left shows this evolution for salt concentrations of 190, 195, 198 and 205 mmol.L$^{-1}$. If one waits for 45 minutes after the salt has been added, we see that for the highest salt concentration, 205 mmol.L$^{-1}$, $G'$ $\simeq$ 4000 Pa, approximately 20 times larger than $G"$. For 198 mmol.L$^{-1}$, $G' \approx G"$, while for 195 mmol.L$^{-1}$, $G"$ is approximately 10 times larger than $G'$, and for 190 mmol.L$^{-1}$, $G'$ reaches zero, to the accuracy of the measurement.
		
	Frequency sweep measurements within the range $10^{-1}$ to $10^2$  Hz  (Fig.~\ref{figRheo}- Center) are performed. For salt concentrations higher than 198 mmol.L$^{-1}$, our materials do not exhibit significant frequency dependence within this interval, with $G'$ always above $G"$. Near the transition, however, there are significant variations of both $G'$ and $G"$ with frequency. For 190 mmol.L$^{-1}$, the colloid is still far from the sol-gel transition, with $G'$ being very low within the whole frequency range and the material behaving like a dilute hard sphere suspension. 

\begin{figure*}
	\centering
	{\includegraphics[width=0.6\textwidth]{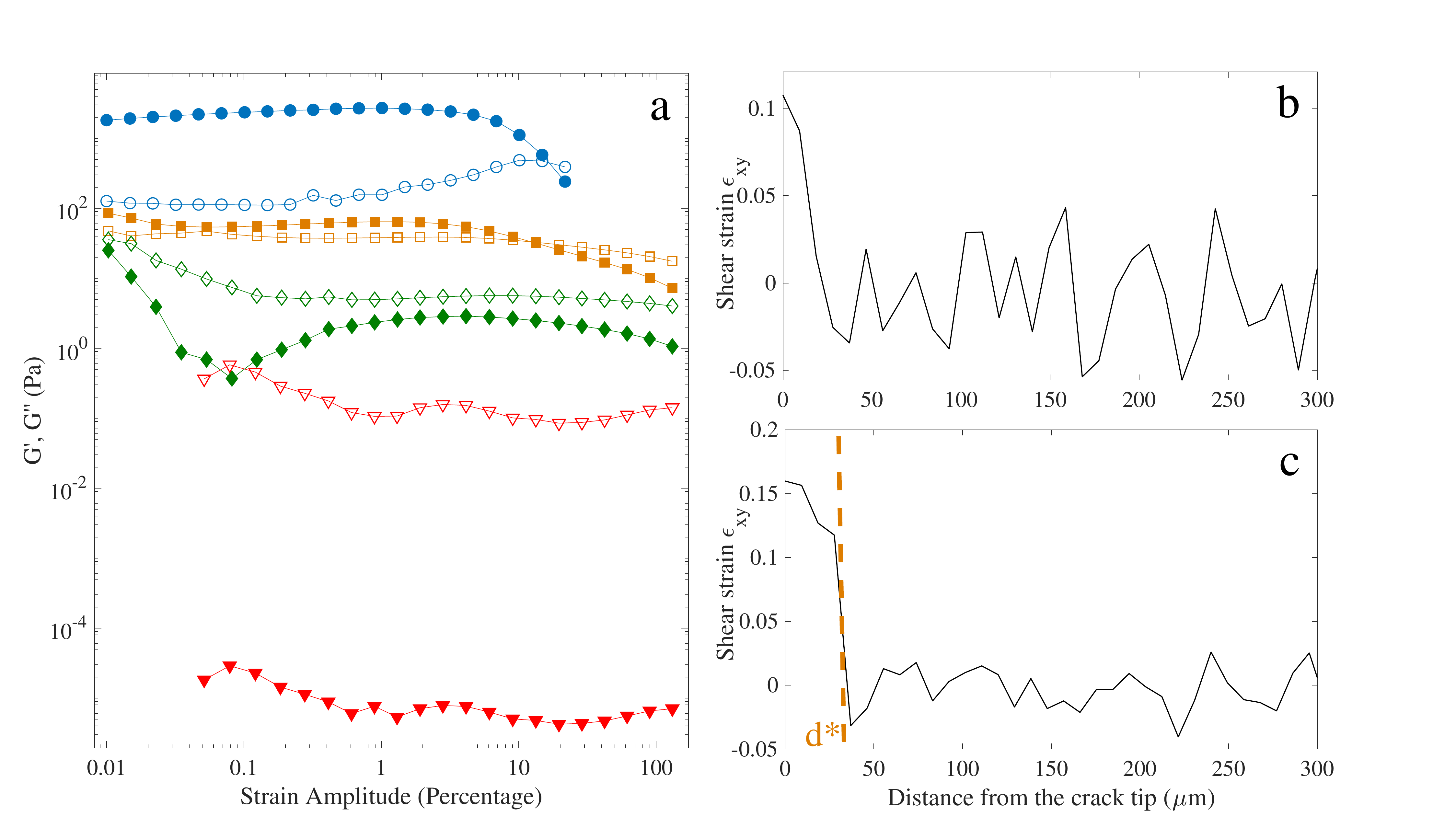}}

	\caption{{(a) Amplitude sweep rheology. $G"$ (empty symbols) and $G'$ (full symbols) as a function of the imposed shear strain amplitude. Red triangles 190 mmol.L$^{-1}$; green rhombus 195 mmol.L$^{-1}$; orange squares 198 mmol.L$^{-1}$; blue circles 205 mmol.L$^{-1}$. At low salt concentration, $G'$ is always smaller than $G''$, while one can observe the emergence of a yield strain $\epsilon_{xy}^Y \simeq$ 15 $\%$ (for 205 mmol.L$^{-1}$) and $\epsilon_{xy}^Y \simeq$ 10 $\%$ (for 198 mmol.L$^{-1}$). (b) and (c) Shear strain $\epsilon_{xy}$ measured with DIC along the $y$ direction, as a function of the distance from the crack tip. (b) Salt concentration 205 mmol.L$^{-1}$. The measured strain is always smaller than $\epsilon_{xy}^Y$. (c)  Salt concentration 198 mmol.L$^{-1}$.  The measured strain is higher than $\epsilon_{xy}^Y$ as long as the distance from the crack tip is smaller than $d^{\star} \simeq$ 37 $\mu$m.}}
\label{figRheoAmplitude}
\end{figure*}

	Amplitude sweep measurements within the range 0.01 to 100~$\%$ (Fig. ~\ref{figRheoAmplitude}a) are also performed. For concentrations lower than 198 mmol.L$^{-1}$, $G'$ is always smaller than $G''$. For higher salt concentrations, the measurements show the existence of a yield strain $\epsilon_{xy}^Y$, above which $G'$ becomes smaller than $G''$. This can be interpreted as a fluidization of the material under stress equivalent to what is found in simulations~\cite{Roy_2016a, Roy_2016b, Johnson:2018aa} and experiments in silica colloidal gels, where the sol-gel transition is controlled either by adding salt like we do~\cite{Persello:94} or by adjusting the temperature~\cite{Colombo_2014}. 

\begin{figure*}
	\centering
	{\includegraphics[width=0.30\textwidth]{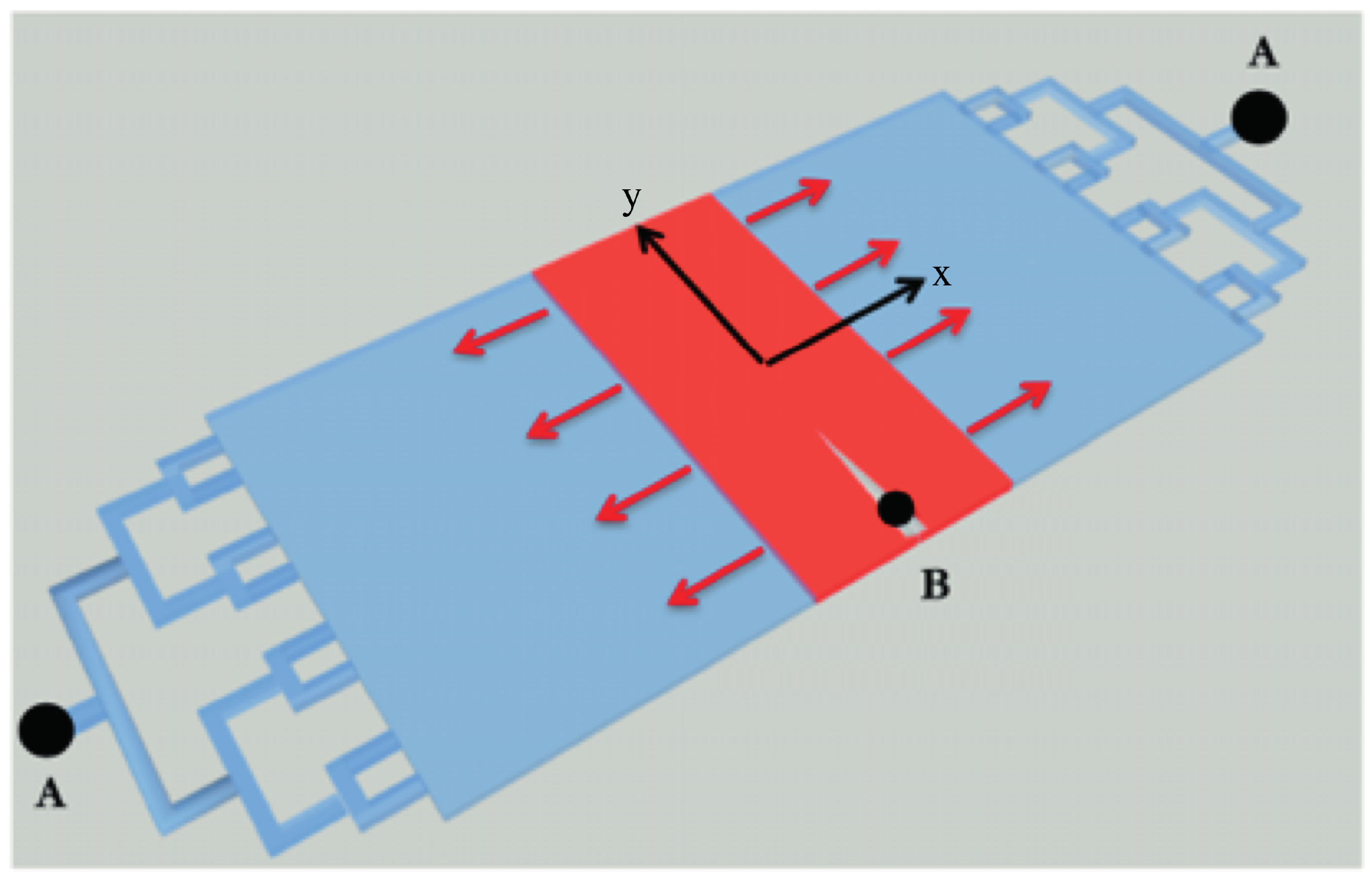}}
	\caption{{Microfluidics-based mechanical test cell. Hele-Shaw chamber containing the sample (red) surrounded by mineral oil (light blue). Oil is sucked out of the chamber from points A, which results in a displacement of the gel boundaries in the $x$-direction. In this plane stress configuration, a crack propagates in the $y$-direction.}}
	
\label{figcell}
\end{figure*}

	We use a setup aimed at fracturing soft materials designed in our group by Lefranc et al.~\cite{Lefranc:14}. It consists of a sealed chamber, with a gap of 350  $\mu$m enclosed by two glass slides and lateral walls made of a  rigid photocurable glue ($NOA  81\textsuperscript{\tiny\textregistered}$, Norland Optical Products). The cell contains an integrated notch of length 5 mm and radius of curvature $\simeq$ 10  $\mu$m. It is fabricated in a clean room using microfluidic stickers technology \cite{Bart:2008}. To ensure perfect slippage and prevent adhesion of the gel samples on the surfaces of the cells, the latter are treated with a solution of 1.5 vol$\%$ of  1H,1H,2H,2H-Perfluorooctyl-trichlorosilane (Sigma-Aldrich) in n-heptane for 15 minutes and then rinsed with isopropanol and dried in a oven at $70~^\circ$ C for 1 hour. A sketch of the cell is shown in Figure~\ref{figcell}.

	Prior to the fracture experiments, the sol is injected into the microfluidic cell at the rate of 1 $\mu$L.s$^{-1}$. This exposes the colloid to a strain rate of no more than $0.5 s^{-1}$ for about one minute. Thanks to grooves in the bottom glass plate, the material stays confined at the center of the cell. The sample is 350 $\mu$m thick and has lateral dimensions 20 mm in the $x$ direction and 10 mm in the $y$ direction.
		
	The empty part of the chamber is then filled with a mineral oil (Light Oil, Sigma-Aldrich), with the entry port being closed with a PDMS plug. The chip is then placed in a reservoir with a saturated atmosphere to prevent drying and put in an oven at $50~^\circ$ C to accelerate gelation. After 45 minutes, the chip is removed from the oven and quenched to room temperature. Note that, since our material is quite sensitive to its thermo-mechanical history, we follow a strict experimental protocol and, in order to avoid any damage of our gels, gelation always take place \textit{in situ} in the microfluidic cell. 
	
	The oil exit ports A are connected to a pair of synchronized syringe pumps (neMESYS Mid-Pressure Syringe Pumps, Cetoni GmbH). Sucking the oil out of the cell at a controlled flow rate $Q$ acts as a low stiffness tensile machine, imposing a displacement to the oil/gel interfaces. Consequently, a single crack is nucleated at the notch and starts to propagate along the $y$ direction (Fig.~\ref{figcell}). This crack will be shown to be very close to a pure mode I (tensile) crack in Section~\ref{secResults}. To prevent elastocapillary effects, the opening B is connected to a water reservoir so that the crack is filled with water throughout the experiment. As a matter of fact, if the surface tension between the filling fluid and the gel is $\gamma $, even perfectly sharp cracks may appear to have a finite radius of curvature of the order of the elastocapillary length:  ${\cal L}=\gamma/E$, where $E$ is the Young modulus of the material. Because we work with water-based colloids, $\gamma $ is negligible in the case of water filling the cracks, while it is equal to $\gamma \simeq$ 7.10$^{-2}$N.m$^{-1}$. Hence ${\cal L}\simeq 70$cm for $E=$1 Pa, and ${\cal L}\simeq 7\mu$m, to be compared with the resolution of the experiment, 1.50$\mu$m for magnification X4, and 0.584$\mu $m for magnification X10. 	  

	For a fixed $Q$, the crack undergoes a brief transient acceleration, and then its speed $V$ varies very slowly as long as the crack is reasonably far from the sample edges. The speed of the observed cracks ranges from 10$\mu $m.s$^{-1}$ to 1cm.s$^{-1}$ as $Q$ is varied from 1 to 500 $\mu$L.min$^{-1}$. Crack propagation can be considered as quasi-static since the speeds are much lower than the minimum Rayleigh wave speed  $V_R \approx $7.5cm.s$^{-1}$, corresponding to the smallest Young modulus, i.e. to the lowest salt concentration of 195 mmol.L$^{-1}$ at which cracks appear.  
	
	The crack tip location is monitored by optical microscopy with 4x and 10x magnifications (4x: field of observation 9 mm$^2$, pixel size 1.5 $\mu$m; 10x: 1.44 mm$^2$, pixel size 0.584  $\mu$m). Images are registered at a frequency of 75 Hz.  The addition of carbon black particles enhances the contrast and therefore enables threshold segmentation of the images. This is necessary in order to get the shape of the crack and locate its tip. Carbon black particles also provide the texture necessary for the measurement of displacement fields with the Digital Image Correlation (DIC) method CorreliQ4~\cite{Hild_06}.
	
	 The displacement fields are measured over the whole field of view, except inside the cracks, which are masked to improve the convergence of the CorreliQ4 algorithm. We use 16x16 pixels$^2$ elements, resulting in a measurement uncertainty of the order of a few centipixels. As the crack is followed during propagation, both the deformed and the reference images correspond to situations under load, taken at sufficiently large time intervals so that the crack tip moves enough (at least 32 pixels) to keep a reasonably low uncertainty level, but at the same time sufficiently small so that the displacement field does not evolve too much.

	The measured displacement field is then projected onto a reference basis of eight different fields given by Linear Elastic Fracture Mechanics (LEFM): the first three correspond to solid body motions (two translations and one rotation), the fourth and fifth are respectively the opening (mode I) and shear (mode II) Irwin asymptotic displacement fields, the sixth one is the displacement field due to the first non-singular stress term in the vicinity of the crack tip (called T-stress), and, finally, the seventh and eighth are higher order subsingular terms. The procedure consists of a minimization of the mean squared differences between the measured displacement fields and the reduced displacement basis (see the post-processing method described in \cite{Roux_2006}). This procedure allows the estimation of the mode I and mode II Stress Intensity Factors $K_{I}$ and $K_{II}$. Let us remind that, in the vicinity of an elastic crack submitted to both a tensile stress $\sigma_{xx}$ and shear stress $\sigma_{xy}$, all the stress components $\sigma_{ij}$ at a distance $r$ from the tip behave in the same way:
\begin{equation}
	\sigma_{ij}= \frac{1}{\sqrt{2\pi r}}(K_I f_{ij}^I(\theta) + K_{II}f_{ij}^{II}(\theta)) +\delta_{yy}Tg(\theta)
\end{equation}	
 where $K_I$ and $K_{II}$ are respectively proportional to $\sigma_{xx}$ and $\sigma_{xy}$.  $\theta$ is the angular coordinate of the considered point, and $f_{ij}^I$, $f_{ij}^{II}$ and $g$ are functions of $\theta$ which depend on the shape of the specimen. $T$ is the T-stress, parallel to the direction of crack propagation. The difference between the projected linear elastic estimation and the original measurement defines a map of residuals, where one can thus locate deviations from linear elasticity.

\section{Results}
\label{secResults}
\begin{figure*}
	\centering
	{\includegraphics[width=0.3\textwidth]{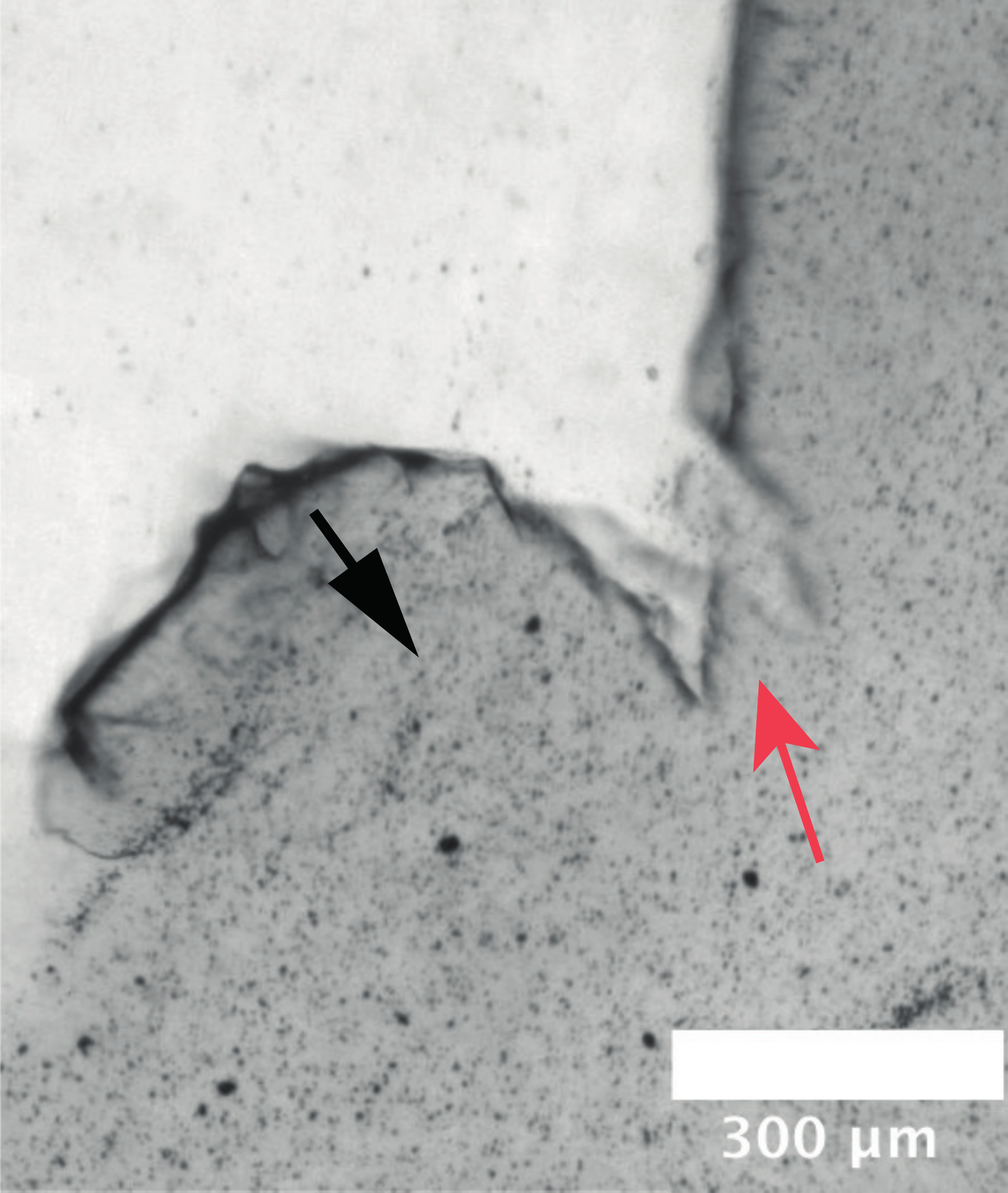}}
	\caption{{Salt concentration 198 mmol.L$^{-1}$. The material starts by flowing, then the front forms a cusp from which two cracks emerge. The black arrow indicates 
	the flow direction, and the red arrow points at the nucleation of two cracks from a cusp of the front.}	
 }
\label{figfingers}
\end{figure*}

For salt concentrations lower than 195 mmol.L$^{-1}$, only flow is observed in the form of  Saffman-Taylor-like viscous fingers~\cite{Saffman}. 
On the contrary, for salt concentrations larger than 202 mmol.L$^{-1}$, a single crack is observed to propagate in each experiment. Both the flow and fracture of the material are observed for salt concentrations between those values. Actually, fingers in this region often give birth to one or even two fissures (see Fig. \ref{figfingers}). This coexistence of flow and fracture will be discussed later. The measurements reported in the following section consider mostly crack behavior. Note that, for each salt concentration, around 12 (resp. 5) experiments are performed for $V=10^{-5}$ m.s$^{-1}$ (resp. $V=10^{-3}$ m.s$^{-1}$). 

We first examine the shapes of the cracks, then use DIC to determine the displacement fields. Both methods provide similar values of stress intensity factors $K_I$ and $K_{II}$, from which we derive the values of the fracture energy release rate. Finally, we show that these experiments exhibit hysteresis.

\subsection{Crack opening displacements}
	
\begin{figure*}
	\centering
	{\includegraphics[width=0.4\textwidth]{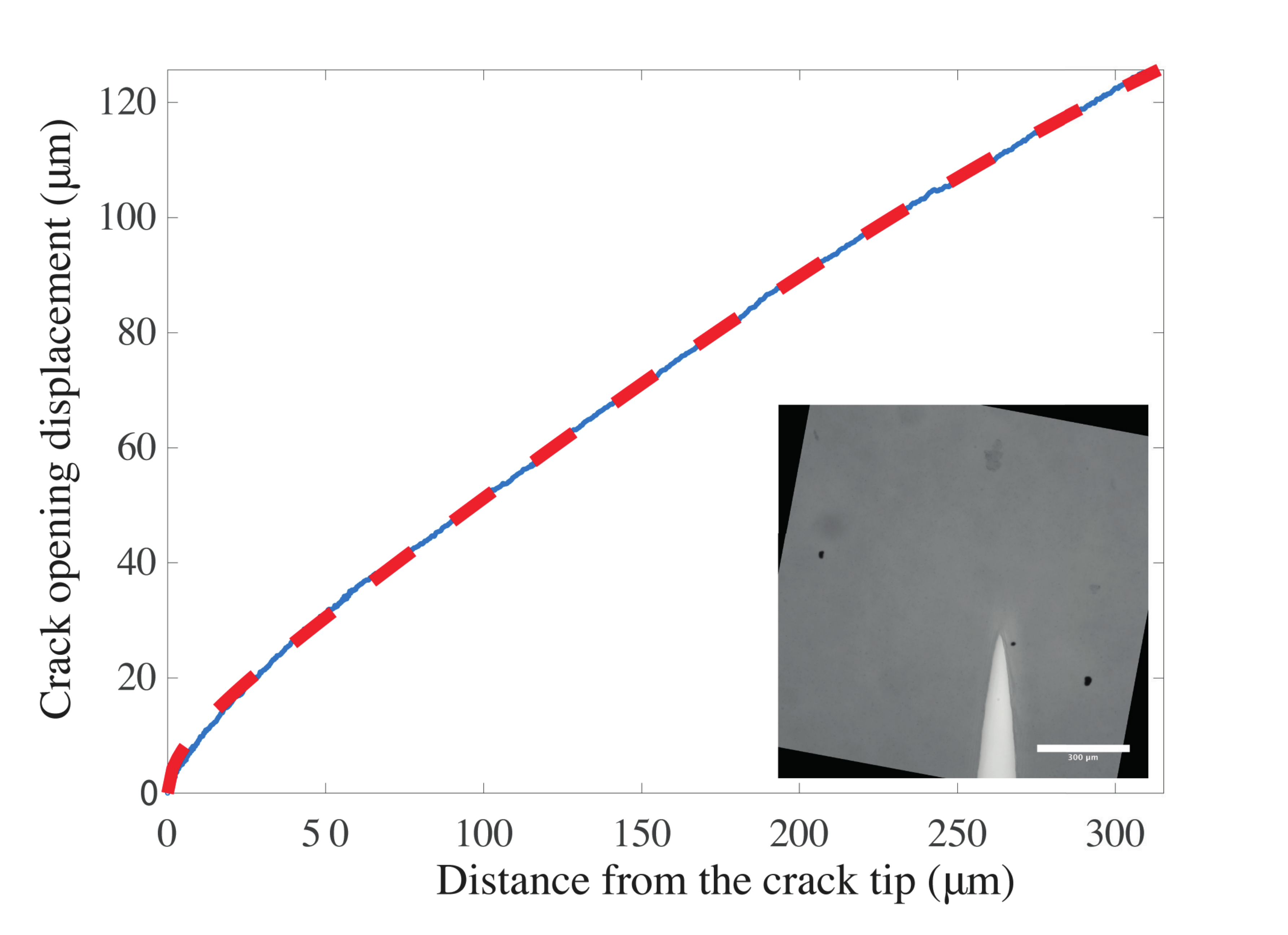}}
	{\includegraphics[width=0.44\textwidth]{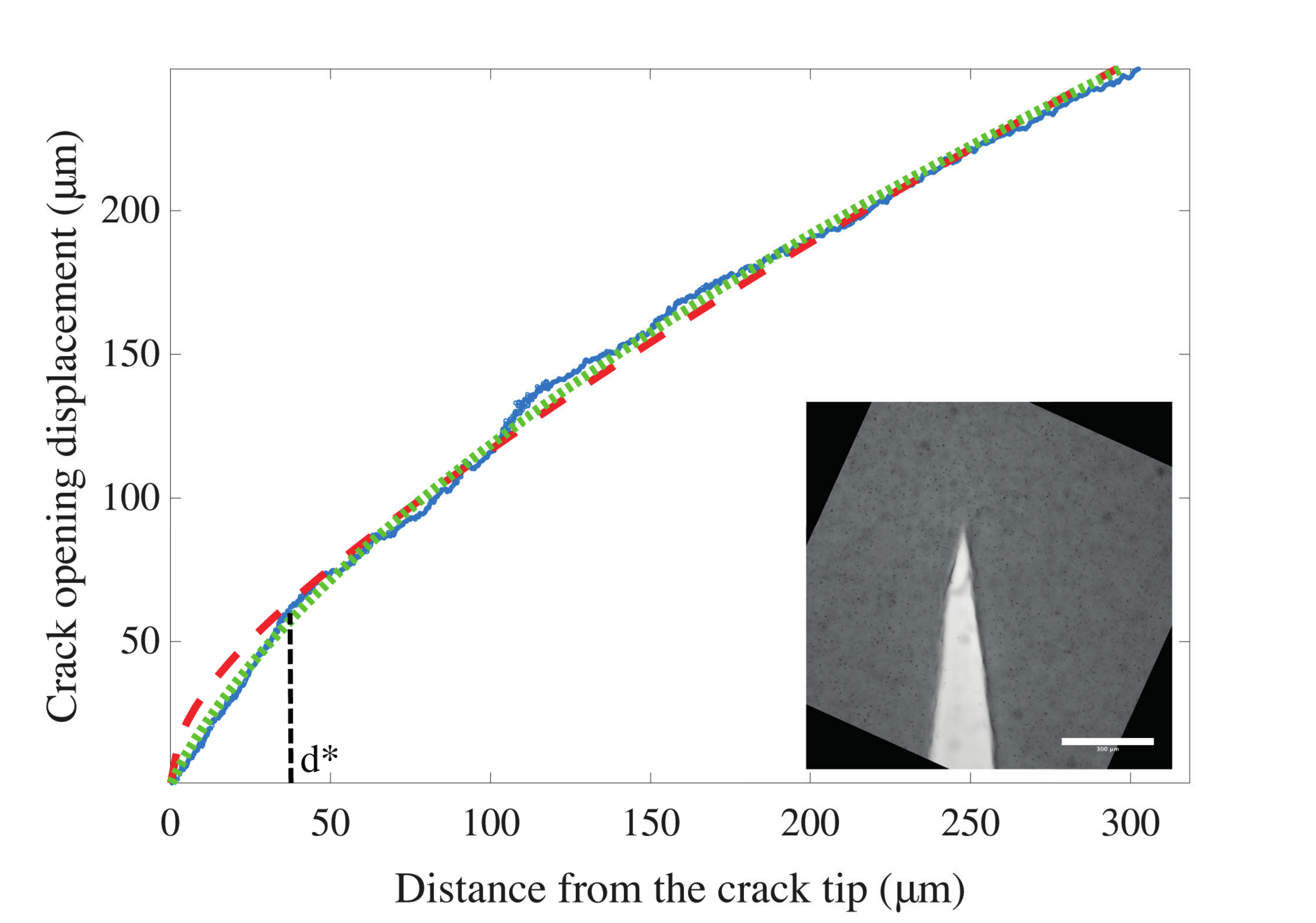}}
	\caption{{{\bf Left}. Salt concentration 205 mmol.L$^{-1}$. The crack edges are smooth, and Williams' linear elastic crack shape, Eq.~\ref{eq:u}, fits it perfectly over the whole region of observation (red dotted line). $K_I\simeq 33$ Pa.m$^{1/2}$. {\bf Right}. Salt concentration 198 mmol.L$^{-1}$. The crack edges are jagged, and it has a linear elastic shape (Eq.~\ref{eq:u}, red dotted line) only at distances  $d>d^{\star}$ from the crack tip. For $d<d^{\star}$, the apex is nearly linear, and the whole profile can be efficiently fitted by Eq.~\ref{eq:umod} (green dotted line). From the large distance elastic shape, one can derive the value of the stress intensity factor (see Eq.~\ref{eq:KI}): $K_I\simeq 0.5$ Pa.m$^{1/2}$. }}
\label{figCracksSlow}
\end{figure*}
	
Figure \ref{figCracksSlow} shows typical cracks progressing in gels with high and low salt concentrations (205 and 198 mmol.L$^{-1}$) at a low velocity 
(10 $\mu$m.s$^{-1}$). Cracks at high salt concentrations exhibit a smooth profile (Fig. \ref{figCracksSlow}a) which is well fitted by the first three terms of the expression predicted by Williams \cite{Williams_57} using Linear Elastic Fracture Mechanics for a pure mode I crack in two dimensions, namely:

\begin{eqnarray}
u(d)={K_I \over E}\sqrt {8\over \pi} d^{1/2}\left[1+ {d\over d_{1}}+ \left( {d\over d_{2}}\right) ^{2} \right]
\label{eq:u}
\end{eqnarray}

\noindent where $u$ is the crack opening displacement, $d$ is the distance from the crack tip, $K_I$ is the mode I crack intensity factor and $E$ is the Young's modulus, and $d_1$, $d_2$ two parameters. In this case, $d_1$ and $d_2$ are respectively found to be 720  $\mu$m and 1 mm, showing that the observed region around the crack tip is mostly dominated by the Irwin term $d^{1/2}$. From this fit, the value of $K_I$ can be determined.
 
\begin{figure*}
	\centering
	{\includegraphics[width=0.6\textwidth]{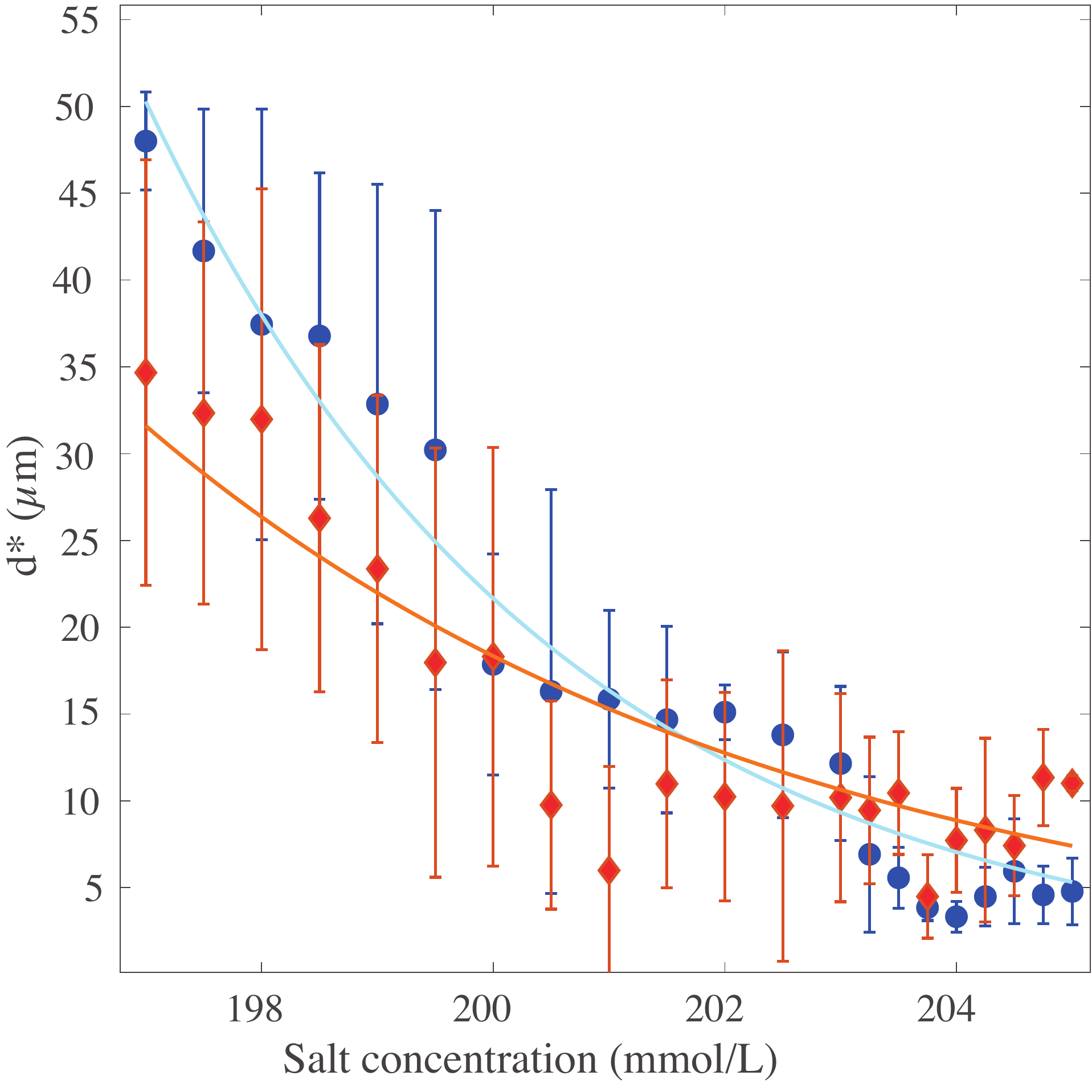}}
	\caption{{ d$^{\star }$ as a function of salt concentration. Blue dots correspond to measurements for $V=10^{-5}$ m.s$^{-1}$. Red dots correspond to  $V=10^{-3}$ m.s$^{-1}$. The two lines are guides-to-the-eye (exponential fits of the low and high velocity data respectively). Error bars in the hysteretic region (salt concentration lower than 200 mmol.L$^{-1}$) are given for the sake of completeness, since flow and fracture actually coexist in this range of salt concentrations.
}}
\label{figdstar}
\end{figure*}

Cracks in gels close to the sol-gel transition are qualitatively different from those at high salt concentration: they are more jagged and depart from a parabolic shape in the vicinity of their tip. As a matter of fact, close to the tip, the measured Crack Opening Displacement (COD, Fig. \ref{figCracksSlow}-b) exhibits a nearly {\it linear} profile: $u(d) \propto d$. To account for this observation in a phenomenological way, we propose the following modified Williams expansion series:
\begin{eqnarray}
\widetilde{u} (d)=d \left[  {{A}\over{1+({d\over{d^{\star}}}) ^{1/2}}}+\left( {d \over \widetilde{d_{1}}}\right) ^{1/2}+\left( {d \over \widetilde{d_{2}}}\right) ^{3/2} \right]
\label{eq:umod}
\end{eqnarray}	
where $d^{\star}$ is the crossover length between the linear apex and the linear elastic crack shape, and $A$ is the slope at the apex. The function $\widetilde{u}$ has the desired limits: it is linear when $d\ll d^{\star}$, and Equation~\eqref{eq:u} is recovered when $d\gg d^{\star}$. Equation \eqref{eq:umod} captures well the shape of the profiles for all salt concentrations. The characteristic lengths $\widetilde{d_{1}}$ and $ \widetilde{d_{2}}$ at which higher order terms  become non-negligible  are larger than 1 mm. The slope $A \approx 2$ does not show any clear dependence, neither on salt concentration nor on crack speed. The salt concentration-dependent mode I stress intensity factor $K_I$ is determined from the large scale fit of the COD:
\begin{eqnarray}
K_I\simeq A\sqrt {\pi \over 8}  \sqrt {d^{\star}} E
\label{eq:KI}
\end{eqnarray}	
Figure \ref{figdstar} shows the evolution of $d^{\star}$ as a function of salt concentration for cracks moving at  $V=10^{-5}$ m.s$^{-1}$ and $V=10^{-3}$ m.s$^{-1}$. We see a continuous increase of $d^{\star}$ as we move from a more cohesive gel, for which $d^{\star}$ is lower than 10  $\mu$m, to a very soft material close to the gel transition, with values of $d^{\star}$ around 50 $\mu$m for slow moving cracks, and 30 $\mu$m for fast ones. 

We also notice a pronounced increase of the dispersion of the measured values of $d^{\star}$ within the salt concentration interval 195-200 mmol.L$^{-1}$. In this region, flow and fracture actual coexist (see Fig. \ref{figfingers}) and small perturbations of the gel time (controlled by the fixed experimental parameters and the salt concentration) can lead to a significant difference in rheology, leading to a large dispersion of $d^{\star}$ for these salt concentrations. 

\subsection{Digital Image Correlation}

\begin{figure*}
	\centering
	{\includegraphics[width=0.3\textwidth]{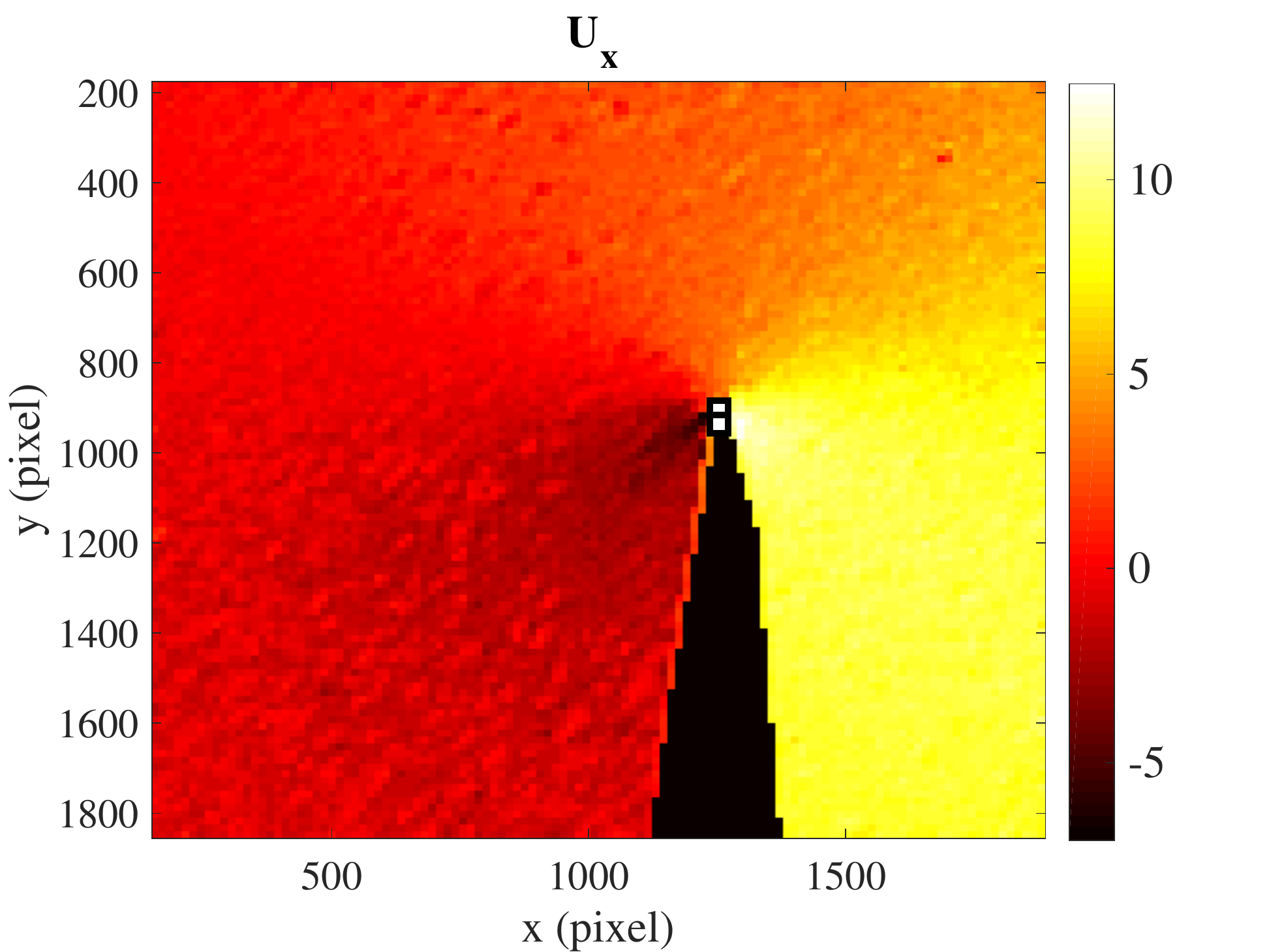}}
	{\includegraphics[width=0.3\textwidth]{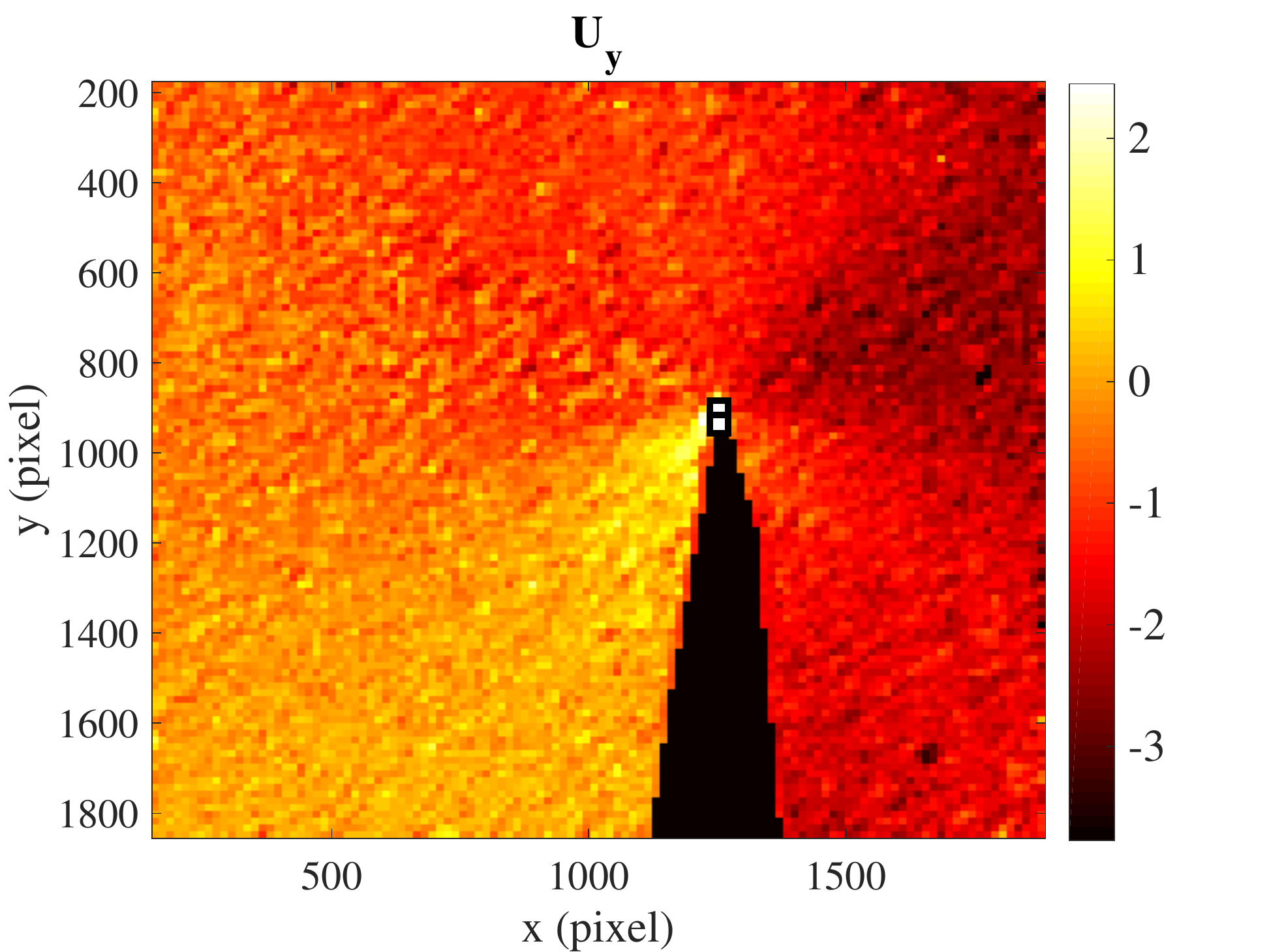}}
	{\includegraphics[width=0.3\textwidth]{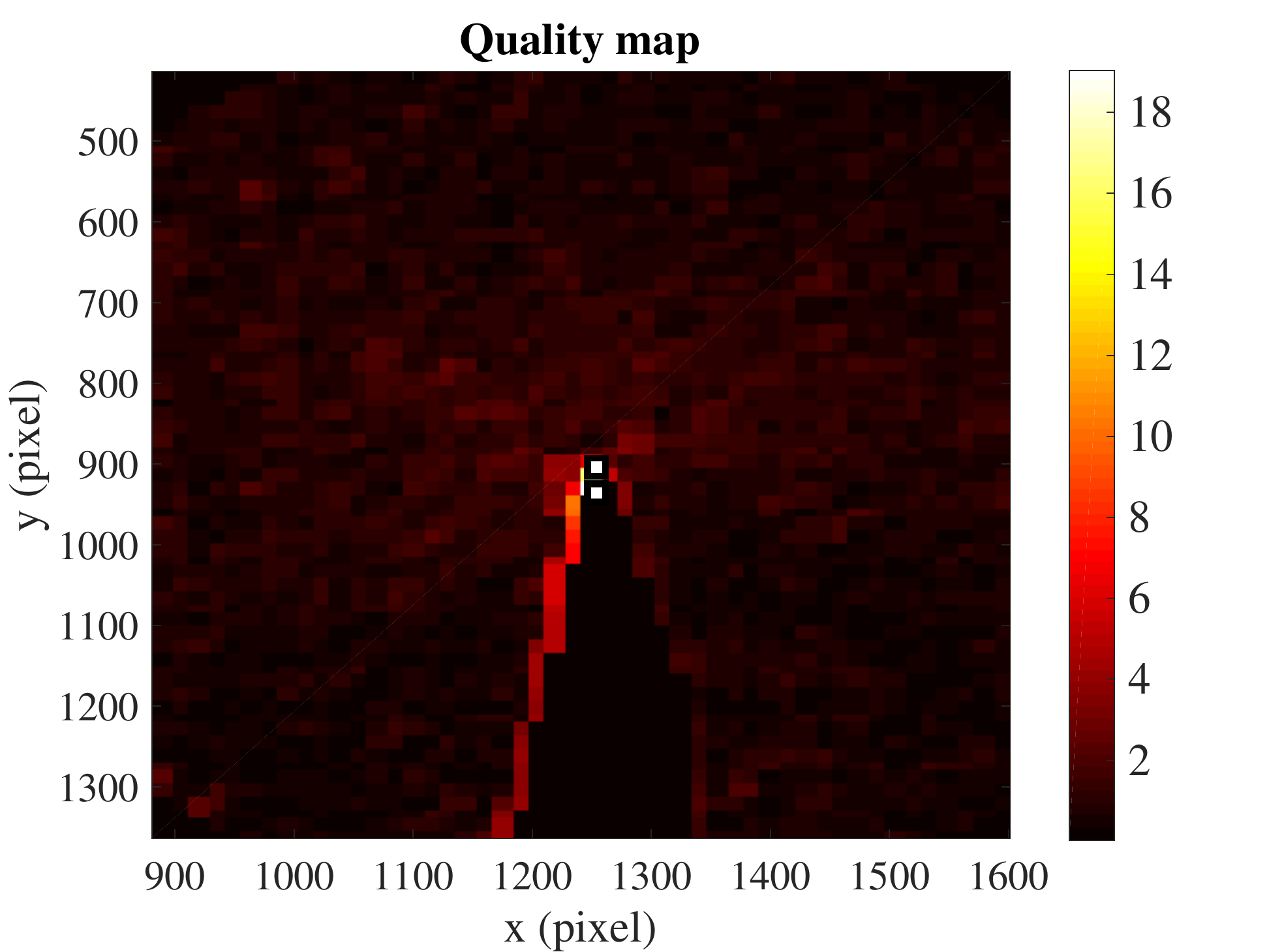}}
		
	{\includegraphics[width=0.3\textwidth]{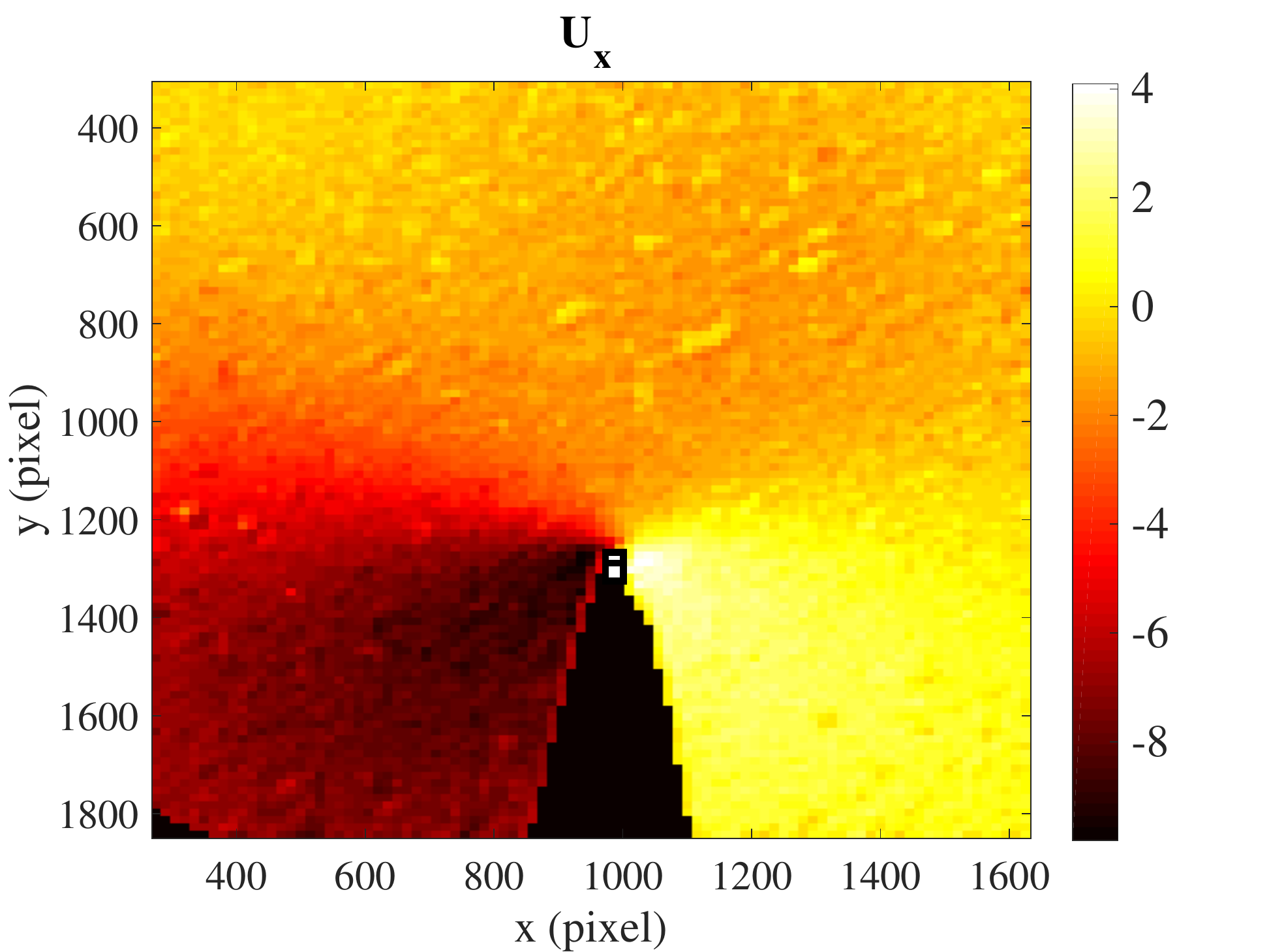}}
	{\includegraphics[width=0.3\textwidth]{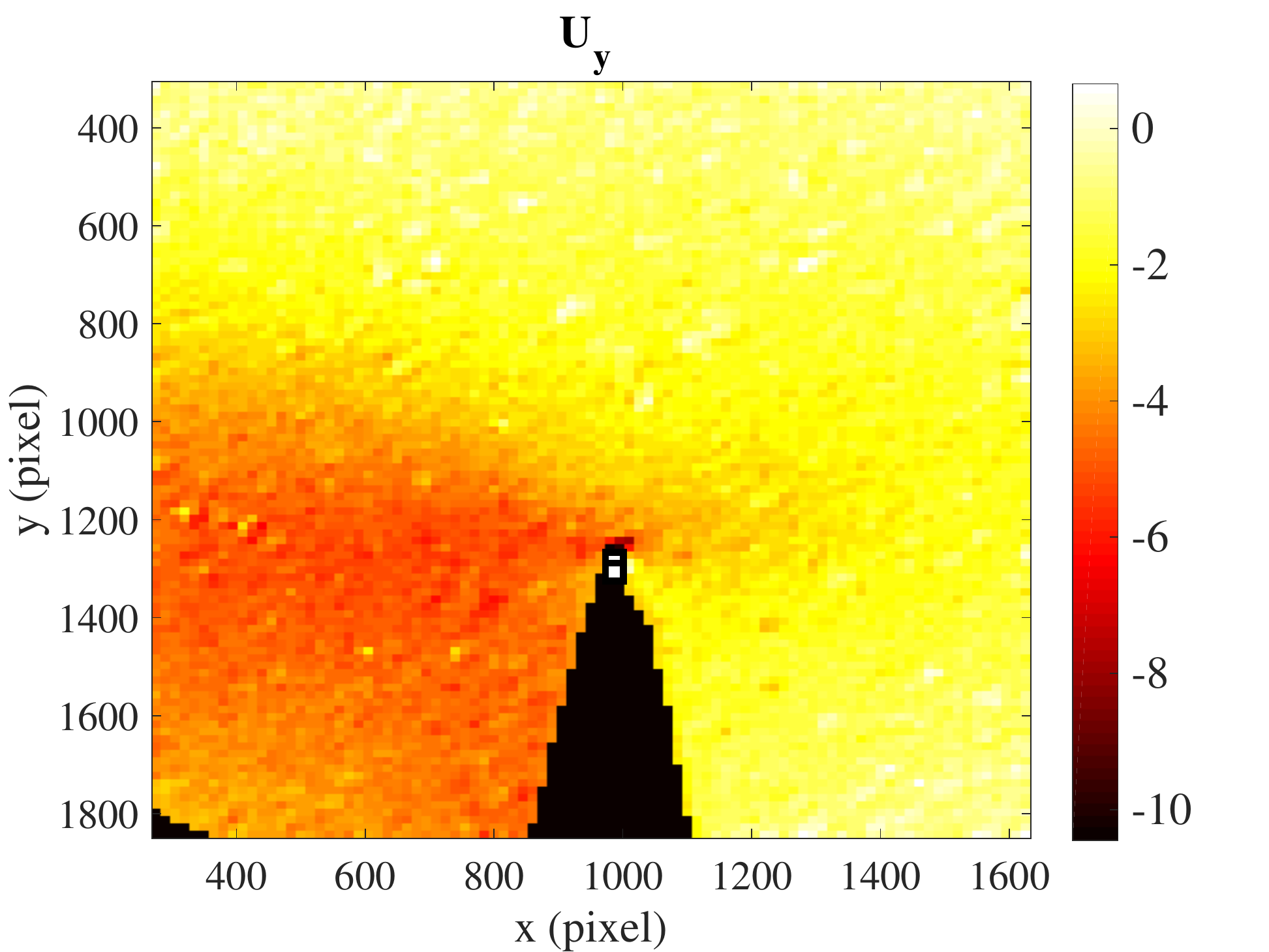}}
	{\includegraphics[width=0.3\textwidth]{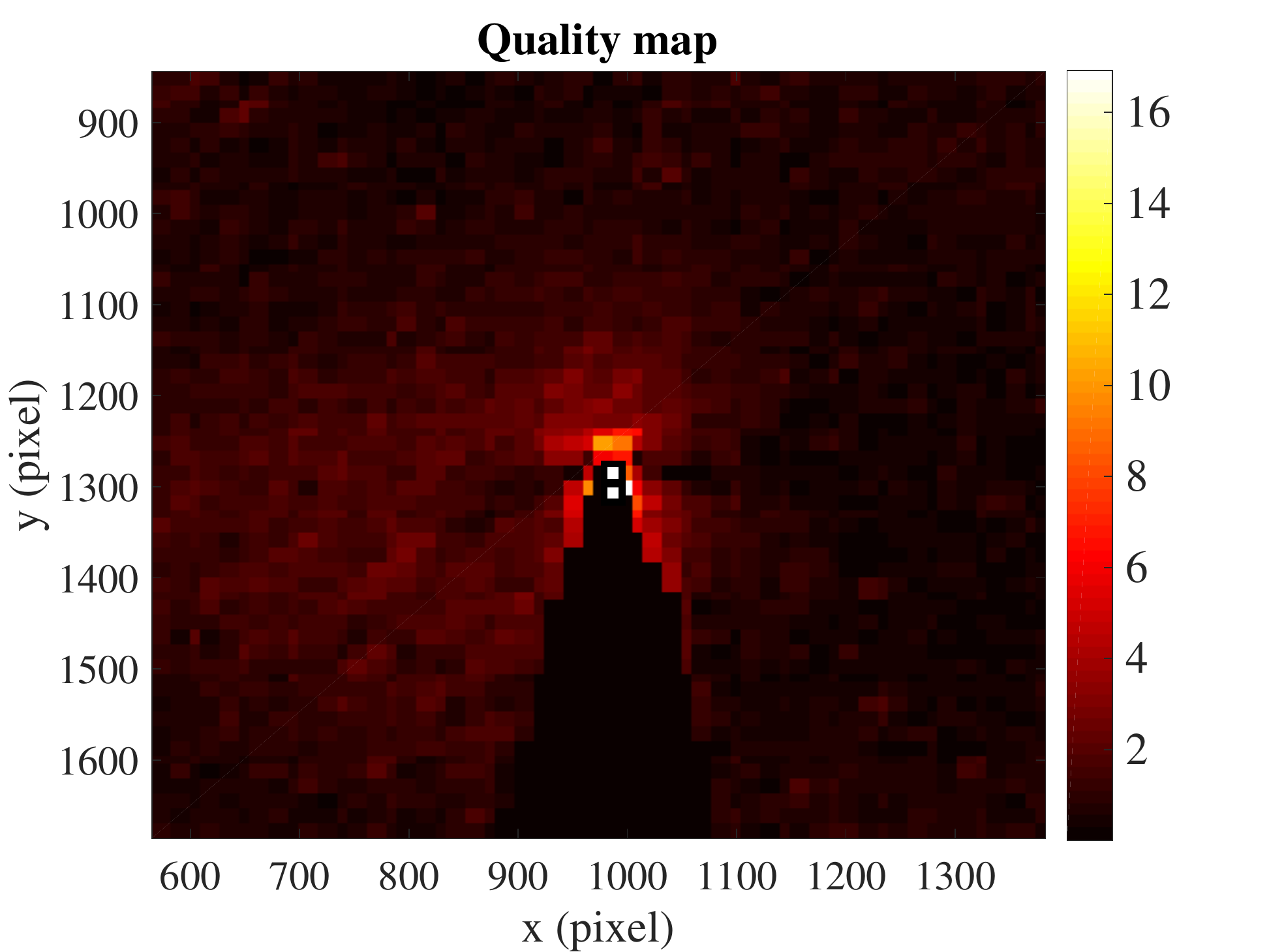}}

	\caption{{Maps of the displacements $U_x$ (direction perpendicular to crack propagation, first column), $U_y$ (direction of crack propagation, second column), and map of residuals (third column) measuring the distance to a perfectly elastic solution. Units are shown in pixels (1 pixel = 0.584  $\mu$m).  The first line refers to a salt concentration 205 mmol.L$^{-1}$. The map of residuals shows that the discrepancy with the elastic solution is high only in the very vicinity of the crack tip. The second line refers to a salt concentration 198 mmol.L$^{-1}$. The map of residuals shows that high differences between the estimated displacement field and the purely elastic solution extend over a region of size $d^{\star }\approx 37 \, \mu$m in this case. Once this region is masked to perform the projection on the elastic basis, the discrepancy is considerably decreased, and one recovers the value of $K_I$ obtained using the shape of the crack.}	
 }
\label{figCorreli}
\end{figure*}

The typical incremental displacement field for slow cracks in cohesive gels is shown in Figure \ref{figCorreli}.  

Concerning high salt concentration materials, the estimated displacement fields shown in Fig.~\ref{figCorreli} correspond quite well to the mode I Irwin displacement fields, except immediately at the crack tip. This is shown in Fig.~\ref{figCorreli} (top line on the right), where the map of the norm of the difference between the measured field and the LEFM one is displayed on the right. Values of the stress intensity factors are $K_I\approx 28.5$ Pa.m$^{1/2}$ and $K_{II}\approx -0.6$ Pa.m$^{1/2}$; T-stress is negligible: $T\approx 0.02$ Pa. 

\begin{figure*}
	\centering
	{\includegraphics[width=0.7\textwidth]{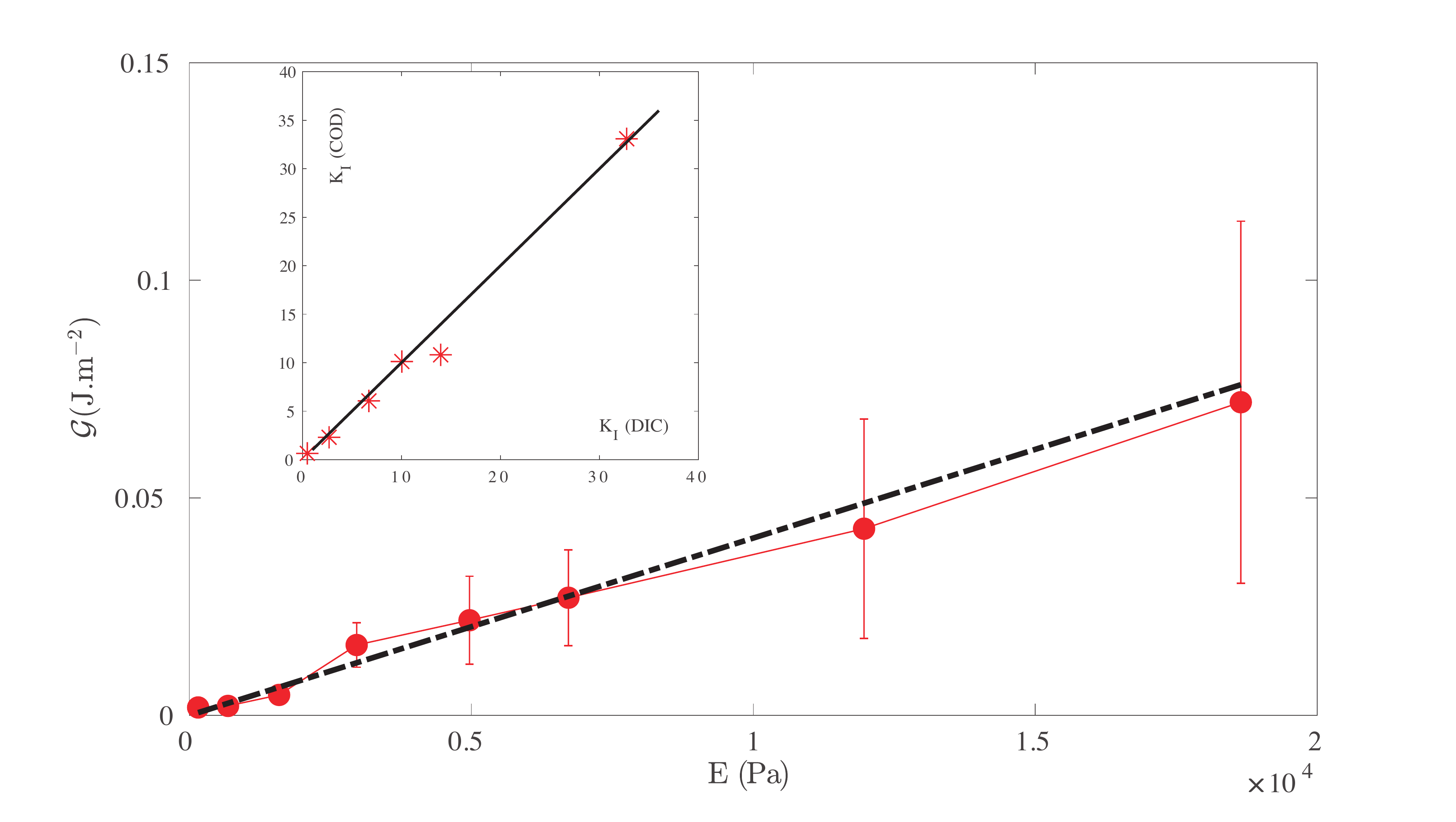}}
	\caption{{ Crack velocity $V$=10$^{-5}$ m.s$^{-1}$. ${\cal G}$ (defined in Eq.~\ref{eq:G}) is plotted against the Young modulus $E$ of the material. {\it Inset} $K_I$ measured from the shape of the crack is plotted against $K_I$ derived from the DIC analysis, showing perfect agreement.}	
 }
\label{G}
\end{figure*}

For low salt concentrations, large discrepancies between DIC displacements and their linear elastic projection are observed in an extended region ahead of the crack tip. To properly recover the Irwin fields, a region of diameter $\simeq d^{\star}$ around the crack tip has to be masked. In this case, masking allows one to estimate the mode I stress intensity factor $K_I\approx 0.32$ Pa.m$^{1/2}$, the mode II stress intensity factor $K_{II}\approx 4.10^{-3}$ Pa.m$^{1/2}$, and the T-stress $T\approx 9.10^{-5}$ Pa. Both $K_{II}$ and $T$ can then be considered negligible, confirming that our experimental setup results in pure mode I cracks. Note that, in all cases, the value of $K_I$ deduced from the DIC analysis after masking is compatible with the $K_I$ obtained from the large distance shape of the crack, as shown in the inset of Fig.~\ref{G}.

Having shown that our cracks are mostly mode I, and after having measured $K_I$, we can estimate the fracture energy given by:
\begin{eqnarray}
{\cal G}= \frac{K_I^2}{E}
\label{eq:G}
\end{eqnarray}
Note that the values of $K_I$ determined from the shape of the crack are remarkably similar to the ones derived from the DIC analysis, as shown in the inset of Fig.~\ref{G}. 
 ${\cal G}$ is plotted as a function of the Young modulus E of the material in Fig.~\ref{G} (see Fig~\ref{figRheo}-Right for the variation of $E$ as a function of the salt concentration). One also notices that ${\cal G}$ (E) does not depend much on the crack velocity $V$.

	We have plotted in Fig.~\ref{figRheoAmplitude}b and c the shear strain $\epsilon_{xy}$, derived from the DIC measurements as a function of the distance from the crack tip. We show that, for a salt concentration of 205 mmol.L$^{-1}$, $\epsilon_{xy}$ is always smaller than the shear yield strain $\epsilon_{xy}^Y$ measured in rheology (Fig.~\ref{figRheoAmplitude}a). In contrast, for a salt concentration of 198 mmol.L$^{-1}$, $\epsilon_{xy}$ is higher than $\epsilon_{xy}^Y$ as long as the distance from the crack tip is smaller than $d^{\star} \simeq$ 37 $\mu$m. Hence, the material is fluidized close to the crack tip: $d^{\star}$ is the distance over which the gel flows under stress.

\subsection{Hysteretic behavior}

\begin{figure*}
	\centering
	{\includegraphics[width=0.4\textwidth]{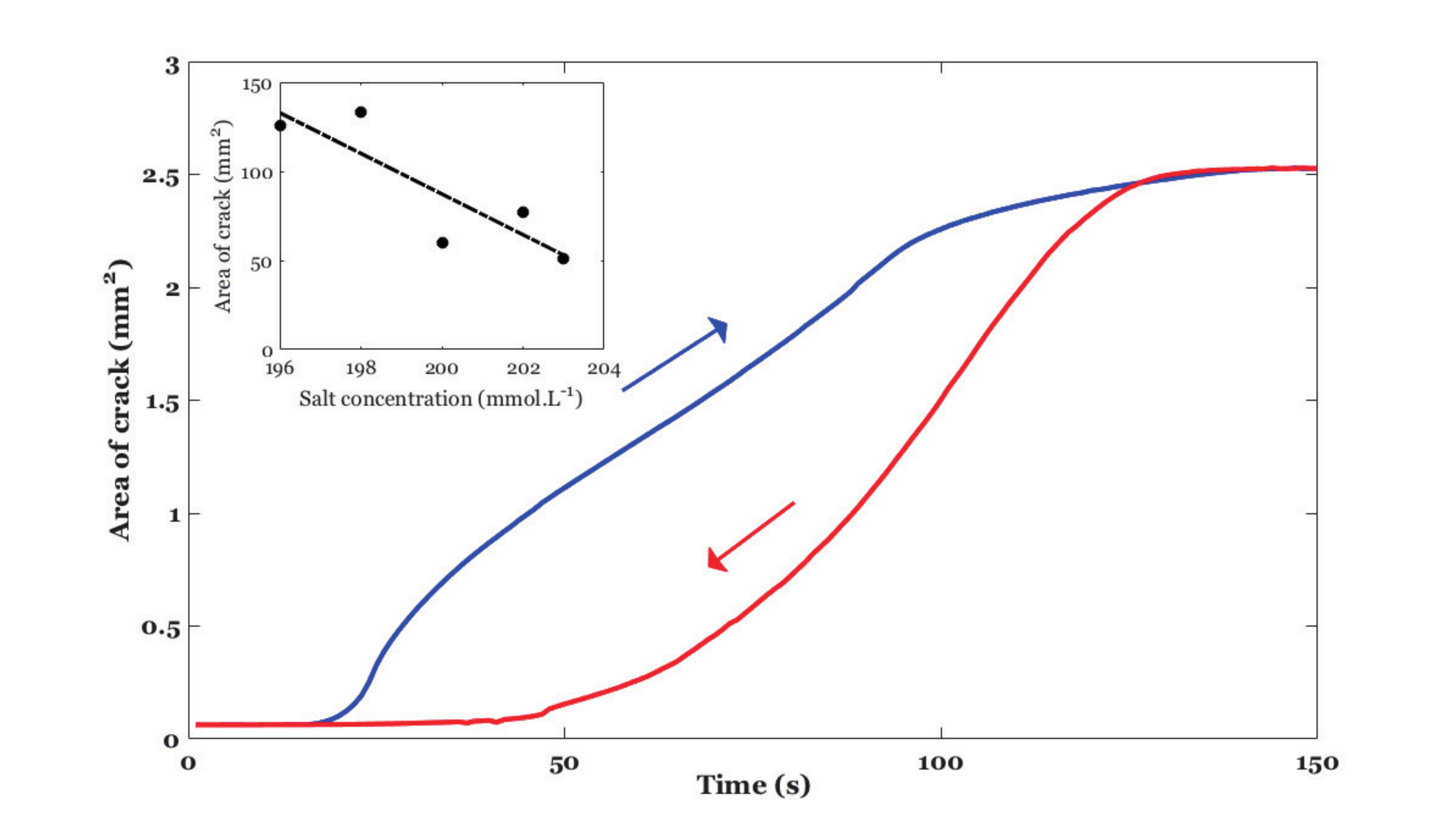}}
	\caption{{Experiment performed for a salt concentration 202 mmol.L$^{-1}$, and a crack filled with air at atmospheric pressure. Area of the crack when it opens (in blue), and when it closes (in red). Opening and closing are performed at a constant oil flow rate 10  $\mu$L/min, inverted from suction to injection at $t = 150$ s. Note that, in this particular case, the elastocapillary length is on the order of ${\cal L}=\gamma/E\approx 70 \, \mu$m, where $\gamma$ is the surface tension. {\bf Inset}. Evolution of the area of the crack with salt concentration. The dashed-dotted line serves as a guide to the eye.}	
 }
 
\label{hysteresis}
\end{figure*}

In the region where coexistence of flow and fracture is observed, we have performed successive fracture and healing experiments. But since closing a crack filled with water is obviously impossible, these experiments have been performed by letting air fill the fissures at constant atmospheric pressure. By sucking out the oil, we open a crack, and by pushing it back, we close it. We then measure the area of the crack (i.e. the volume occupied by the air inside the crack, divided by the thickness of the sample), and see that it is significantly larger when it opens than when it closes, as seen in Fig.~\ref{hysteresis}.

\section{Discussion}

We have analyzed how flow morphs into fracture in the vicinity of a sol-gel transition. We observe that there is a region close to the transition (salt concentrations 190-195 mmol.L$^{-1}$) where the material only flows as a viscous fluid and cracks are never observed. In an intermediate domain of salt concentrations (195-202 mmol.L$^{-1}$), we observe both fracture and flow. Furthermore, the phenomenon is hysteretic in that region: an opening crack is systematically wider than a closing one.

We have seen that, when the material is far in the gel phase, cracks are very close to being elastic at all observable scales. Closer to the sol-gel transition, cracks are only elastic at large distances from their tip, i.e. at distances larger than a characteristic length $d^{\star}$. $d^{\star}$ can be understood as the distance over which the material flows around the crack tip. As a matter of fact, we have shown that the material fluidizes over a scale $d^{\star}$ around the crack tip: the high shear strain induced by the presence of the loaded crack exceeds the yield strain above which $G''$ becomes larger than $G'$. Poroelasticity~\cite{Lopatnikov10_TranspPorousMed,Baumberger_EurPhysJE06} may play a role despite the slow time scales at which it occurs, but its contribution is difficult to pinpoint. This flow at the crack tip can be observed directly for salt concentrations as low as 196 mmol.L$^{-1}$ (see the video in Supplementary Materials). 

	The trace of this flow is at the origin of the jagged shape of cracks, as noticed above. $d^{\star }$ grows up to $\approx$48 $\mu$m for a salt concentration of 195 mmol.L$^{-1}$ when the crack velocity is $V=10^{-5}$ m.s$^{-1}$, and only to $\approx$ 37 $\mu$m when the crack velocity is higher, $V=10^{-3}$ m.s$^{-1}$. This is typically a viscoelastic effect. 

For salt concentrations ranging from 190 to 195 mmol.L$^{-1}$, no crack is observed, which can be seen as a $d^{\star}$ being on the order of the entire crack length. Although it seems extremely difficult practically, varying the sample size within an extended range would allow one to decide whether, for a system of infinite size, $d^{\star}$ actually diverges or remains finite but large. As a matter of fact, both the coexistence of flow and fracture and the hysteretic behavior in the intermediate region corresponding to salt concentrations 195-202 mmol.L$^{-1}$ indicate that the flow to fracture transition is akin to a first-order transition. The gelation of silica nanoparticles at low to intermediate volume fraction by an increase of the ionic strength, such as our system, is typically understood as a Reaction Limited Colloidal Aggregation of irreversible nature \cite{Cao:10, Schantz2006, Kolb96}. Under these conditions, the colloidal sol to gel transition is currently described as a continuous geometric transition more akin to a second order process \cite{Kolb96, Abete06}, contrasting to what we observe in terms of its flow and fracture behavior.

One could expect to observe a growing viscous energy dissipation as $d^{\star}$ increases, i.e. as the salt concentration decreases. On the contrary, we observe that ${\cal G}$ is simply proportional to the Young modulus $E$, and can thus be interpreted as the energy per unit area needed to create two free surfaces in the material~\cite{Joanny79}. This is actually the classical Griffith energy term, which does not include any dissipation. This may appear surprising, since all our other measurements show that the material undergoes viscous flow over a growing distance $d^{\star }$. In fact, the viscous dissipated energy turns out to be negligible compared to the surface energy. To understand this, let us compare the power $P_s$ spent in surface creation during a time $\delta t$, when the crack grows by $V\delta t$, with the viscous dissipated power $P_v$. Clearly, one has $P_s=e_sVb$, where $b$ is the thickness of the sample, and $e_s$ the surface energy per unit area. Neglecting the strain ($\epsilon _{ij}$) gradients within the region of volume $V^2 \delta t^2 b$, one can infer: 
\begin{eqnarray}
P_v=2\eta \frac {\epsilon _{ij} \epsilon _{ij}}{\delta t^2}V^2 \delta t^2 b
\label{eq:Pv}
\end{eqnarray}
so that one can find an upper bound to the ratio $P_v/P_s$:
\begin{eqnarray}
\frac {P_v}{P_s} < 2\eta \frac {(\epsilon _{ij} \epsilon _{ij})_{\text{max}}V}{e_s}
\label{eq:Pv2}
\end{eqnarray}

This ratio is expected to be higher at low salt concentrations, and for the highest velocity. Our experiments for 196 mmol.L$^{-1}$ (lowest salt concentration at which we observe cracks) show as largest strains of $\approx$ 0.2. Hence, with $\eta \approx 0.1$ Pa.s, $V=$10$^{-3}$ m.s$^{-1}$ and $e_s\approx $6.10$^{-4}$ J.m$^{-2}$, we find that the upper bound of the ratio $P_v/P_s$ is 7.10$^{-3}$. The viscous dissipated energy is therefore indeed negligible with respect to the surface energy. As a consequence, from the point of view of the fracture energy release rate, one can say that all our materials, whatever their distance to the sol-gel transition, behave in a perfectly elastic way. In other words, the properties of the observed cracks are only modified at a short distance by the flow of the material: it does not affect the long-distance properties, as for example the mode I stress intensity factor.

As far as the detailed shape of the cracks close to their tips is concerned, one would need to couple compression or large deformations rheology experiments with Finite Elements simulations, as it was done for an agar gel in ~\cite{Rong_ExtMechLet16}. However, we know that colloidal gels typically follow an exponential hardening constitutive law~\cite{Gisler99}, which is in agreement with the observed quasi-linear apex.

Finally, it should be noted that we have not been able to access the microscopic mechanisms at stake in our system, because of the very small size of the silica beads. We are currently working on other colloidal gels for which the basic entity has a size of $\sim $1 micrometer~\cite{gim18}, and microemulsion-based materials with elementary size $\sim $50 micrometers.

In conclusion, we have shown that the transition from flow to fracture through the second order sol-gel transition may well be a first order phase transition. It involves the growth of a length scale which characterizes a distance over which the material fluidizes close to the crack tip. Other systems where the sol-gel transition is controlled in a different way should be explored as well. It would be particularly interesting to couple this experiment with the analysis of local structural modifications. We also believe that this experiment is able to shed a different light over less well understood liquid-to-solid transitions, such as the glass transition~\cite{DynHet11,Berthier_Physics11,Albert_Science18}. In the future, it could also help research on the toughening of gels~\cite{Zhao_2014, Zhang16}.

\section*{Acknowledgements} The authors acknowledge funding by the French Agence Nationale de la Recherche (ANR-14-CE05-0037-01). Microfabrication was performed thanks to the Technological Platform of the Institut Pierre-Gilles de Gennes (CNRS UMS 3750). They are also indebted to F. Hild for enlightening discussions, and to J.-P. Bouchaud for a careful reading of their manuscript and very interesting suggestions. 
\balance


\bibliography{ManuscriptGIMENES-Bibliography} 
\bibliographystyle{rsc}

\end{document}